\documentstyle[12pt,aaspp4,epsf]{article}

\date{}
\def\fun#1#2{\lower3.6pt\vbox{\baselineskip0pt\lineskip.9pt
  \ialign{$\mathsurround=0pt#1\hfil##\hfil$\crcr#2\crcr\sim\crcr}}}

\def\kms{${\rm km\, s^{-1}}$\space}
\def\msun{${M_{\odot}}$\space}
\def\msunyr{$M_{\odot}${\rm{yr}}$^{-1}$\space}
\def\lsun{${L_{\odot}}$\space}

\def\ha{$\rm H{\alpha}\;$}
\def\h2{$\rm H_2\,$}
\def\ergss{erg\,s$^{-1}$\space}

\def\hn~{ H$\alpha$+[\ion{N}{2}]\space}
\def\n2{[\ion{N}{2}]}

\def\beq{\begin{eqnarray}}
\def\eeq{\end{eqnarray}}

\def\bc{\begin{center}}
\def\ec{\end{center}}

\singlespace

\slugcomment{To Appear in The Astronomical Journal, October 1996}

\lefthead{ Valluri \& Anupama}
\righthead{H$\alpha$ Imaging of HCG 62}

\begin{document}

\title{H$\alpha$ Imaging of the Hickson Compact Group 62\footnote[1]{Based
on observations using Vainu Bappu Telescope, VBO, Kavalur, India.}}

\author{M. Valluri}
\affil{Columbia University, Department of Astronomy,\\
 New York, New York, 10027\\}
\and
\author{G. C. Anupama}
\affil{Indian Institute of Astrophysics,\\ Bangalore 560034, India}

\begin{abstract}

We report results of a search for optical emission line gas in the
Hickson compact group 62. From narrow band (\ha) imaging observations
we estimate $10^4$~\msun of ionized hydrogen gas within the central
regions of three of the galaxies in the group.  There has been
considerable interest in this group since {\it ROSAT} detected an
extended halo of diffuse X-ray emission, possibly with a cooling flow
at the center. The optical emission line luminosities from the three
galaxies are comparable to each other ($\sim 10^{39}$~\ergss) and
resemble those seen in normal X-ray luminous early type
galaxies. There is no evidence for any excess emission in the galaxy
which lies at the center of the X-ray cooling flow. We find that the
most likely source of excitation for the emission line nebulosities is
photoionization by young stars.

\end{abstract}


\section{Introduction}

Compact groups are dense concentrations of galaxies with densities
comparable to those in the centers of rich clusters. They have always
been of interest since their high galaxy densities and small crossing
times make them ideal sites for studying galaxy interactions and
mergers and the products of these processes.  In the last few years
{\it ROSAT} observations of several groups of galaxies have shown that
compact groups are frequently associated with extended diffuse X-ray
halos. The first {\it ROSAT} images of extended diffuse X-ray halos of
hot (1~keV) gas in the compact group HCG62 (Ponman \& Bertram 1993)
and the group NGC 2300 (Mulchaey et al. 1993) imply that in both
systems the group galaxies are embedded in a giant dark matter
halo. Since these first observations at least 8 other groups with
diffuse X-ray emission arising from the extended hot gas have been
observed (Ebeling, Voges \& B\"ohringer 1994 and Pildis, Bregman \&
Evrard 1995a).  An interesting feature of all the groups with extended
X-ray emission is that they are dominated by early type galaxies.

In the group HCG62 (Ponman \& Bertram 1993, PB hereafter) the X-ray
surface brightness peaks on the dominant galaxy HCG62$a$ and decreases
outwards.  PB found that the spectral hardness varied significantly
with radius and fitted multiple Raymond-Smith hot plasma models to the
X-ray spectrum in annular rings to show that the gas temperature is
low at the very center rises sharply outwards to peak at about
3$^{\prime}$ from the center and then decreases again towards larger
radii. Pildis et al. (1995a) estimate a similar radial temperature
profile beyond 3$^{\prime}$.  Both sets of authors estimate a central
temperature of $\sim 0.8$~keV for the gas. For this temperature, PB
estimate a cooling time for the gas in the central region of $9\times
10^8 h_{50}^{-1/2}$yr. The drop in central temperature and the
presence of a high surface brightness central component suggests the
presence of a cooling flow in this system.

From the earliest X-ray observations of clusters (Lea et~al.~1973) and
subsequent theoretical considerations (Cowie \& Binney 1977, Fabian \&
Nulsen 1977) it has been known that since the cooling time for the hot
gas in many clusters is shorter than a Hubble time, under hydrostatic
equilibrium conditions the cooler gas will sink to the center of the
potential under the weight of the overlying hot gas. Fabian \& Nulsen
(1977) predicted that the cooler components of the resulting ``cooling
flow'' would be detectable in optical emission lines as the gas cooled
through $10^4$K. Extensive observations of the X-ray emission in
clusters have shown that the gas in cooling flows is inhomogenous and
consists of gas of varying temperatures and densities throughout the
cooling radius (100--200~kpc).  Optical line-emitting nebulae have
been observed at centers of about 40\% of central cluster galaxies
(Cowie et~al.~1983, Hu et~al.~1985, Johnstone et~al.~1987, Edge
et~al. 1992) and in 50--60\% of normal early type galaxies
(Demoulin-Ulrich et~al.~1984, Phillips et~al.~1986).  In central
cluster galaxies the optical emission lines arise from asymmetrically
distributed bright filamentary structures generally within the central
10-20~kpc regions of the galaxy. These regions have LINER spectra
(Heckman~et~al.~1989, hereafter HBvBM), with line widths of $\sim
100$~\kms due to turbulent velocities, suggesting that a possible
source of excitation could be low velocity shocks or a dilute
power-law continuum.  The ratios of the [\ion{S}{2}]$\lambda\lambda
6717,6731$ lines imply that the electron densities are of the order
$10^3$~cm$^{-3}$ (HBvBM).  The presence of the [\ion{O}{1}]$\lambda
6300$ line indicates that the emission lines arise from semi-ionized
regions, possibly the ionized-outer skins of cooler clouds (Baum
1992). The strongest piece of evidence connecting the hot gas to the
emission line regions is the fact that emission-line nebulae are only
detected in those clusters with cooling times shorter than
$10^{10}$~yr (Hu et~al. 1985).  While this is evidence that
emission-line nebulae and cooling flows are associated, short cooling
times and large X-ray derived mass accretion rates do not appear to be
a sufficient condition for the existence of emission-line regions
(e.g. A2029, Edge et al. 1992). X-ray cooling flows with optical
emission line nebulae generally also have an excess of UV/blue
continuum emission (Johnstone~et~al.~1987, McNamara \& O'Connell~1989,
Allen~et~al.~1992) frequently correlated with the emission-line
luminosities, distinguishing them from giant elliptical galaxies where
such blue continuum emission is rarely observed.

Over the last decade extensive optical spectroscopy (e.g. Phillips
et~al.\ 1986 [PJDSB hereafter], Kim 1989) and narrow band imaging of
early type galaxies (e.g. Demoulin-Ulrich et al.~1984, Shields 1991)
have established that these galaxies have a significant mass
($10^2-10^4$\msun) of warm ($10^4$K) ionized gas and dust.  This warm
ionized component is detected in emission lines with luminosities of
$10^{38}-10^{40}$~\ergss\, \, in the \hn~ lines alone. Both the narrow
band imaging surveys and spectroscopic studies indicate that the
ionized emission line component is generally in a disk about 1-1.5~kpc
in radius probably rotating about the nucleus.  The gas kinematics
indicates that gas disk has a rapidly rising inner rotation curve with
peak rotation velocities frequently as high as 100~\kms (Kim 1989). In
addition the gas is generally more centrally peaked than the
background continuum and is not aligned with the continuum image of
the host triaxial elliptical galaxy.  This lack of correlation between
the continuum image and the orientation of the gas disk could imply
that the gas and stars are kinematically and dynamically decoupled. In
some cases the kinematics of the gas clearly suggests that cold gas
was acquired through an accretion or merger event with a gas rich
late-type companion galaxy was subsequently ionized by the hot gas
(Macchetto \& Sparks 1992).  The existence of dust lanes supports this
hypothesis in several galaxies (e.g. Goudfrooij et al. 1994). As in
the case of cluster nebulae these nebulae generally have LINER spectra
(although \ion{H}{2} region like spectra are observed in a few cases
[e.g. PJDSB]). Gas densities are similar to those in cluster nebulae
(100-1000~cm$^{-3}$).  Emission line luminosities are found to
correlate with the absolute B magnitudes of the galaxies (PJDSB,
Macchetto et~al.~1996) and systems with radio sources are found to
have more luminous emission line nebulae (Buson et al. 1993, hereafter
B93). Whether a connection exists between the X-ray fluxes and the
emission-line fluxes is even less certain than in the case of cluster
cooling flow nebulae.

The observation of a possible X-ray cooling flow in the compact group
HCG62 provides a situation that is intermediate between X-ray cooling
flows in clusters and those in individual X-ray luminous galaxies. The
central cooling time for the X-ray gas in this system is $\sim
10^9$~yr and the X-ray images suggest that the cooling flow is
centered on the galaxy HCG62$a$.  We imaged the compact group HCG62 to
search for warm ionized gas associated either with a global cooling
flow or with the individual galaxies. In this paper we present images
of the \hn~ emission-line regions in three of the galaxies in this
group.
 
In \S~2 we describe the observations and the data reduction. In \S~3
we present the main results of the observations. In \S~4 we discuss
possible models to account for the observed emission line fluxes. In
\S~5 we summarise the main conclusions from this paper.

\section{Observations and Data Reduction}

The compact group HCG62 was observed using the 2.3~m Vainu Bappu
Telescope at Kavalur, India, on 1994 March 24 and 25. The CCD camera
system was used at the prime focus. This system consists of a GEC back
illuminated chip, with $385\times 578$ pixels of $23\times 23\,\mu m$
size. The image scale is $0.56\times 0.56$~arcsec. The characteristics
of the CCD system are described in Prabhu, Mayya \& Anupama (1992) and
Anupama et al.~(1993).  The seeing was $\sim 2.5$~arcsec on both
nights.

\begin{table}[thb] 
\begin{center}
\caption{ Journal of observations}
\begin{tabular}{rcr}\\ \hline \hline
Date (UT) & Filter & Exposure \\
1994 March &       & \qquad \qquad sec \qquad \qquad \\ \hline \\
24.892 \qquad & $V$ & 600 \quad \qquad\\
24.906 \qquad & $V$ & 1200 \quad \qquad\\
24.810 \qquad & $R$ & 600 \quad \qquad\\
24.818 \qquad & $R$ & 300 \quad \qquad\\
24.842 \qquad & H$\alpha$ & 1800 \quad \qquad\\
24.866 \qquad & H$\alpha$ & 1800 \quad \qquad\\
25.849 \qquad & H$\alpha$ & 1800 \quad \qquad\\
\hline
\hline
\end{tabular}
\end{center}
\end{table}

Seven images were obtained, three narrow band
($\Delta\lambda=80$~\AA), centered at $\lambda 6680$~\AA\
corresponding to the H$\alpha$ emission redshifted to $z=0.0137$, the
mean redshift of the group (Hickson 1993), and two images each in the
broadband $V$ and $R$ filters.  The details of the observations are
given in Table 1.

All frames were reduced following standard procedures.  Each frame was
bias-subtracted using bias values from the overscan region, and
flat-field corrected with averaged twilight flats taken through the
corresponding filter. The bias subtracted and flat-field corrected
images were then partly cleaned for cosmic ray hits. Sky subtraction
was effected using a sky value measured in those parts clearly
unaffected by the light from the galaxies and stars in the field. The
frames were then aligned using positions of the stars in the field and
a shift in both the $X$ and $Y$ directions. Individual frames in each
of the filters were then co-added. To obtain a continuum-free
narrow-band image a rescaled $R$ frame was subtracted from the narrow
band \hn~ image.  The rescaling of the continuum image is generally
the most tricky proceedure and in this case it was estimated using the
stars in the field. In addition to the stars being completely
subtracted from the resultant images, the outer regions of the
galaxies were also fully subtracted out, giving us greater confidence
in the scale factor. The IRAF software package was used for
reductions.

The spectrophotometric standard G60-54 (Oke 1990) was observed through
$V$, $R$ and the narrow band filter and used for flux calibration. The
observed magnitude of the standard star in the \hn~ filter was
compared with the standard magnitude (Oke 1990), and the zero point
correction estimated.  This zero point correction was applied to the
\hn~ fluxes estimated for the individual galaxies.  The fluxes were
corrected for atmospheric extinction determined as $10^{-2.5\,X\,
B/\lambda^4}$, where $\lambda$ is in $\mu$m, $X$ is the airmass and
$B=0.014$~mag $\mu$m$^4$ is the mean differential extinction
coefficient for the site.  Thin clouds were present during both nights
resulting in errors in the magnitudes of 0.1~mag in H$\alpha$ and
0.07~mag in $R$. These translate to an error of 15\% in the H$\alpha$
flux estimates from the images, including all other systematic errors.

\section{Results}

\begin{figure}[thb]
\begin{center}
\epsscale{0.6}
\plotone{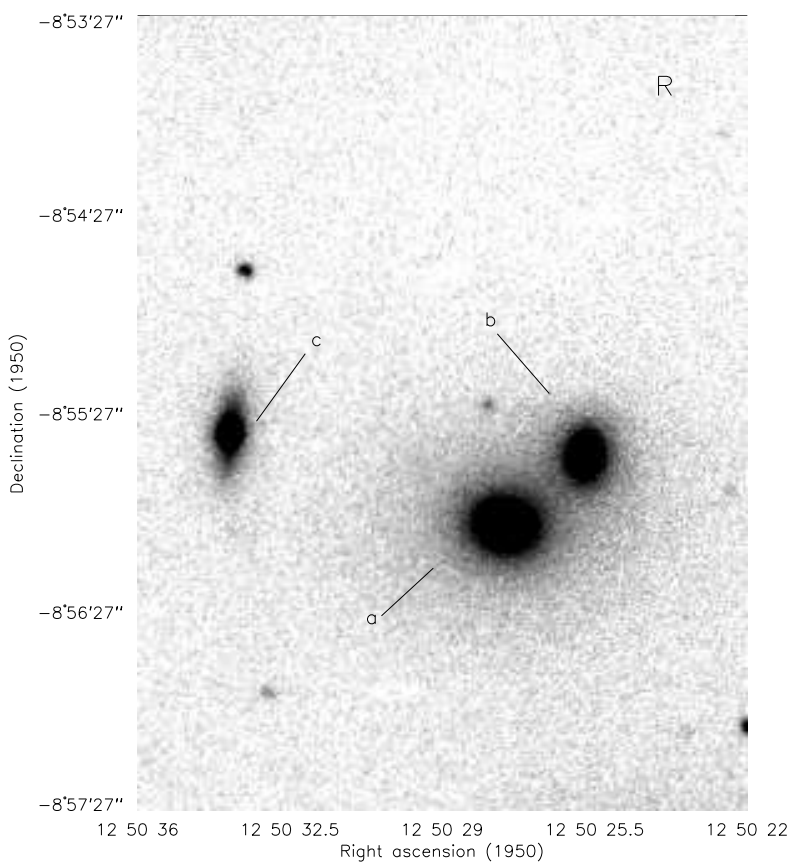}
\figcaption{\label{fig:rband} Coadded {\it R} band image with galaxies
 {\it a, b, c} as marked. }
\end{center}
\end{figure}

The coadded $R$ band image of the group is shown in 
Figure \ref{fig:rband}  with the
galaxies marked {\it a, b, c} as in the Hickson catalogue (Hickson
1982).  HCG62{\it d} was not observed. The observed galaxies are
classified as E3 (HCG62$a$), and S0 (HCG62$b$ and HCG62$c$) (Hickson
1982).  In deep images the galaxies $a$ and $b$ overlap significantly
and are embedded in a diffuse common stellar envelope. Pildis et
al. (1995b) find this diffuse envelope to be roughly symmetric about
galaxy $a$, with some enhancements in the direction of galaxies $b$
and $c$. This diffuse envelope could imply that the central galaxy is
the product of a previous merger.  No other signs of interaction or
merger, such as tidal tails, are visible even in deep images (Pildis
et al. 1995b).

\subsection{H$\alpha$ Fluxes}

\begin{figure}[thb]
\begin{center}
\epsscale{0.6}
\plotone{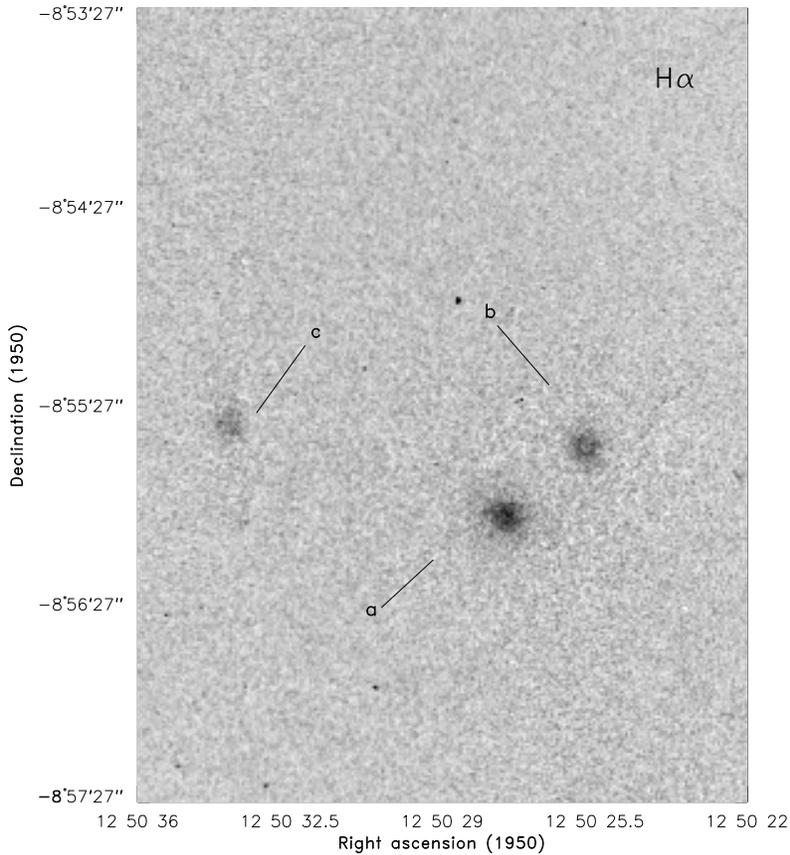}
\figcaption{\label{fig:halpha}Continuum subtracted emission-line image.} 
\end{center}
\end{figure}

Figure \ref{fig:halpha} shows the continuum subtracted
\hn~ line-emission from the
compact group HCG62. Emission is detected from the central regions of
the three group galaxies observed.  The total extent of the
emission-line region is $17$~arcsec in HCG62$a$, $13$~arcsec in
HCG62$b$ and $11$~arcsec in HCG62$c$, assuming a mean distance to the
group of 83~Mpc~{\footnote{In this paper all distance dependent
quantites assume $H_0 = 50$~\kms{$\rm Mpc^{-1}$} and $q_0 = 0$}} (PB)
the corresponding diameters are 6.8~kpc, 5.2~kpc and 4.4~kpc
respectively.  In the galaxy HCG62$b$ the emission appears to arise in
a partial annular ring. This feature is detected within the
point-spread-function (PSF) of the frame, and could be a result of
slight differences in the PSF in the \hn~ and $R$ frames. A similar
feature is however not detected in either HCG62$a$ or HCG62$c$
indicating that the annular distribution in HCG62$b$ could be real.
Annular emission regions have occassionally been detected in the past
in X-ray bright elliptical galaxies (e.g. NGC~4636 [B93] and NGC~3607
[Singh et al. 1994]). Imaging under better observing conditions are
required before this feature can be established.

The total emission-line flux from each galaxy was obtained using
apertures of radius 8.5~arcsec, 6.6~arcsec and 5.5~arcsec,
respectively, for the galaxies HCG62$a$, HCG62$b$ and HCG62$c$.  In
Table 2 we list the observed \hn~ fluxes and luminosities and the \ha
~fluxes and luminosities which are estimated by assuming that the
ratio of [\ion{N}{2}]/\ha = 1.38 (PJDSB).  The observed luminosities
of $\sim 10^{39}$ ergs~s$^{-1}$ are comparable to those observed in
the emission line regions of luminous elliptical galaxies (e.g.
PJDSB, Kim 1989).

\begin{table}[thb]
\caption{H$\alpha$+[N~I\negthinspace I] Fluxes}
\begin{center}
\begin{tabular}{cccccc} \hline \hline \\
Galaxy & \multicolumn{2}{c}{flux/$10^{-15}$}
 & \multicolumn{2}{c}{luminosity/$10^{39}$ } &$\log({\rm 
F_{H\alpha+[N~II]}/F_B})$ \\
       & \multicolumn{2}{c}{ergs cm$^{-2}$ s$^{-1}$}
       & \multicolumn{2}{c}{ergs s$^{-1}$} & \\
\cline{2-6}\\
       & \hn~ & H$\alpha^a$ & \hn~ & H$\alpha^a$ & \\
\hline\\
HCG62$a$ & 1.91 & 0.803 & 1.57 & 0.66 & -4.10 \\
HCG62$b$ & 1.25 & 0.525 & 1.03 & 0.43 & -4.06 \\
HCG62$c$ & 0.93 & 0.391 & 0.77 & 0.32 & -3.86 \\ 
\hline
\hline\\
\multicolumn{6}{l}{$^a$ H$\alpha$ fluxes corrected for \n2 using
\n2/H$\alpha$ = 1.38 (PJDSB)}\\
\end{tabular}
\end{center}
\end{table}

\begin{figure}[thb]
\epsscale{0.8}
\plotfiddle{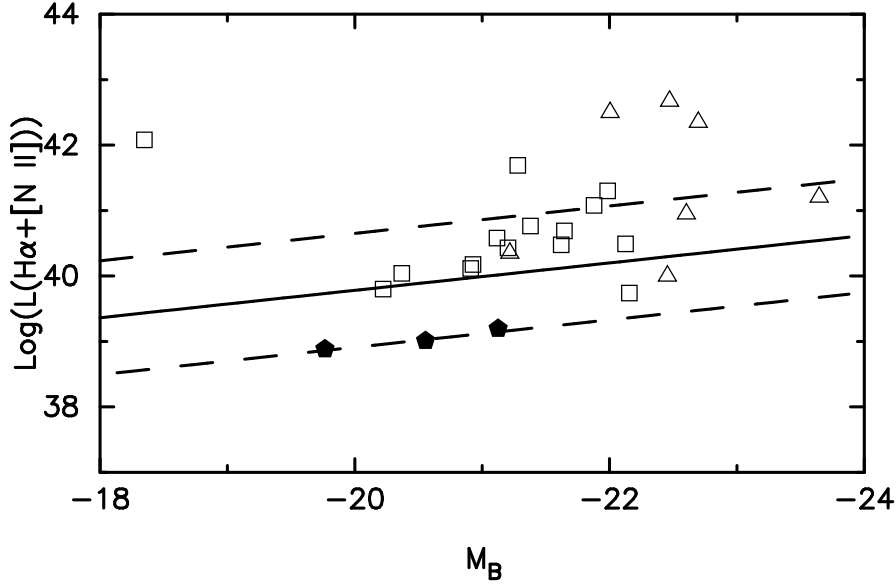}{8cm}{0}{80}{80}{-252}{-72}
\figcaption{\label{fig:pjdsb} Best fit line (solid) with
$1\sigma$ errors (dashed) for the emission-line luminosity versus
absolute B magnitudes for magnitude-limited sample of early type
galaxies from PJDSB.  HCG62 galaxies (filled pentagons), luminous
emission-line nebulae in elliptical galaxies (open squares) (B93) and
cooling flow galaxies (open triangles) (Heckman et al. 1989).}
\end{figure}

It has been known for some time that the emission line luminosities of
normal early type galaxies are correlated with their absolute
magnitudes, and that luminous ellipticals are more likely to have
emission lines than smaller galaxies (PJDSB).  In cooling flow
clusters the emission line luminosity of the central galaxy is found
to correlate with the X-ray inferred mass accretion rate (Johnstone et
al. 1987). Furthermore, HBvBM showed that cooling flow nebulae have
emission line luminosities well in excess of the emission-line
luminosity of a typical giant elliptical galaxy.  Figure \ref{fig:pjdsb}
 shows the
total \hn~ emission-line luminosity for the HCG 62 galaxies (filled
pentagons) plotted against absolute magnitude. The solid line is a
linear fit (with one sigma errors) to the emission-line luminosity
versus absolute B-magnitude (not including upper limits) for a
complete magnitude limited survey of early-type galaxies (PJDSB).
(The fit obtained by B93 {\it includes} upper-limits and has a larger
slope and a smaller intercept but comparable 1-$\sigma$
errors). Elliptical galaxies with luminous emission-line nebulae
(generally associated with nuclear radio sources) (open squares) from
B93 and cluster cooling flow galaxies (open triangles) from HBvBM are
also plotted in the figure.  This figure clearly shows that the
emission-line nebulae in the three HCG62 galaxies are not as luminous
in emission lines for their optical magnitudes as are the luminous
elliptical galaxies or the cooling flow galaxies.  We estimate the
quantity $\log({\rm F_{([N~I\negthinspace I]+H\alpha)}/F_B})$, the
ratio of fluxes in emission-lines and B-band for each of the galaxies
$a, b,$ and $c$ (column 6 Table 2) taking the B magnitudes of these
galaxies as (13.47, 14.04, and 14.83 respectively [from
NED]).{\footnote{ The NASA/IPAC Extragalactic Database (NED) is
operated by the Jet Propulsion Laboratory, California Institute of
Technology, under contract with the National Aeronautics and Space
Administration.}}  This quantity was also computed for the emission
line and cooling flow galaxies: for the B93 sample the mean was
$-2.67\pm 0.86$ and for the sample of HBvBM the mean was $-1.55\pm 2$.
Thus the estimated flux ratios for the HGC62 group galaxies (Table 2)
are between 1--2 standard deviations {\it below} the mean value of the
B93 sample and the HBvBM sample.  While this deviation is not
statistically significant enough to claim that these galaxies do not
belong to the sample of highly luminous emission-line/cooling flow
objects it seems unlikely that they are cooling flow nebular systems.
Furthermore the variation between the individual objects is
signifcantly smaller than the standard deviation of the galaxies in
either the B93 sample or the HBvBM sample, indicating that the group
dominant galaxy which is at the focus of the X-ray cooling flow is no
more luminous in emission lines (relative to B band continuum) than
its companions. We therefore conclude that the observed emission line
nebulae resemble those in normal elliptical galaxies and are not
necessarily the result of accretion from the cooling flow associated
with the group. It may be noted here that a similar ratio of emission
line to B-band flux is also observed in NGC 2300 (Goudfrooij et
al. 1994).

An absence of excess emission from the dominant galaxy is rather
surprising since the X-ray image suggests that the cooling flow is
centered on this galaxy.  It is also surprising since the galaxy
HCG62$a$ hosts a moderate nuclear radio source ($\log(P/{\rm W
Hz^{-1}}) = 20.22$ at 1635 MHz) (Menon \& Hickson 1985) and a
correlation exists between the existence of highly luminous emission
line nebulae and nuclear radio sources (HBvBM, Shields 1991, B93). It
must be emphasised however that while nearly 60\% of early type
galaxies in the field have emission line regions (B93) only about 39\%
of X-ray selected cooling flow cluster galaxies have detectable
optical emission line regions (Edge et al. 1992, Donahue et al. 1992).
This suggests that rather special conditions are required to power the
emission line regions in cluster cooling flows.

\subsection{Mass of Ionized Gas}

The mass of the warm ionized gas component can be estimated from the
H$\alpha$ luminosity ($L_{{\rm H}\alpha}$). Under Case B recombination
conditions, we have \begin{eqnarray} L_{{\rm H}\alpha} = V \epsilon
[\alpha_H^{\rm{eff}}h\nu_{H\alpha}]n_e n_{\rm{H}} ,
\end{eqnarray} where $V$ is the volume of the emitting region,
$\epsilon$ is the filling factor and $\alpha^{\rm{eff}}_{{\rm
H}\alpha}$ is the effective recombination coefficient for H$\alpha$
emissivity (Osterbrock 1989), $n_e$ is the electron density and $n_H$
is the hydrogen density.  Assuming that the gas is fully ionized and
$n_p = n_{\rm H}=n_e$ we estimate the mass of ionized hydrogen in each 
of the three galaxies (Table 3) by the
expression, \begin{eqnarray} {{M_{\rm{gas}}}\over{M_{\odot}}} =
2.8\times 10^2\biggl({{D}\over{10{\rm Mpc}}} \biggr)^2\times
\biggl({{F(H{\alpha})}\over{10^{-14}{\rm ergs\, s^{-1} cm^{-2}}}}
\biggr)\biggl({{10^3 {\rm cm^{-3}}}\over{n_e}}\biggr) \, \, ,
\end{eqnarray} (Kim 1989) where $D$ is the distance to the source 
and $F({\rm H}\alpha)$ is the H$\alpha$ flux
corrected for contribution from the \n2 line using the mean
\n2/H$\alpha$ ratio of 1.38 (HBvBM, PJDSB).  The
estimated mass depends on an accurate estimate of the electron
density in the nebula.  Densities measured directly from
spectroscopic data using the [S~II] line ratios (approximately unity)
for a sample of early type galaxies (PJDSB) imply typical values for
the electron density of $n_e \sim 10^{3}$~cm$^{-3}$.  For cluster
cooling flow galaxies HBvBM estimate densities in the
range 150--400~cm$^{-3}$ and derive filling factors of
10$^{-6}-10^{-7}$.  Alternatively if we assume that the hot and warm
components are in pressure equilibrium then $n_X T_X = n_e T_H$
(where $T_H$ is the temperature of the emission line nebula). With
$n_X = 0.1$ (typical X-ray estimated density for the hot gas in the
central region), $T_H = 10^4$~K we estimate an electron density of
$n_e = 100$~cm$^{-3}$ for $T_X =10^7$~K, the estimated temperature
for the gas in the central region (PB).  In Table 3 we list the
estimated masses of ionized hydrogen  for three possible values of
$n_e$ using H$\alpha$ fluxes in the equation above.

\begin{table}[thb]
\caption{Electron density and hydrogen mass}
\begin{center}
\begin{tabular}{lccc} \hline \hline\\
  Galaxy & {$n_e = 100$ cm$^{-3}$} &{$n_e = 400$cm$^{-3}$}& {$n_e = 10^3$cm$^{-3}$} \\
 \cline{2-4} \\

       & {$M_{\rm{H}}$} ($10^4$~\msun) & {$M_{\rm{H}}$} ($10^4$~\msun) & {$M_{\rm{H}}$} ($10^4$~\msun) \\
\hline \\
HCG62$a$ & 1.52 & 0.39 & 0.15 \\
HCG62$b$ & 1.01 & 0.25 & 0.10 \\
HGC62$c$ & 0.83 & 0.19 & 0.08 \\
\hline
\hline\\
\end{tabular}
\end{center}
\end{table}

To estimate the volume filling factor of the gas we assume that it
occupies a spherical volume of radius determined by the extent of the
emission.  The filling factor are given by
$\epsilon=V_{\rm{gas}}/V_{\rm{tot}}$, where $V_{\rm{gas}} =
M_{\rm{gas}}/n_e m_{\rm{H}}$, and $V_{\rm{tot}}={{4}\over{3}}\pi r^3$
(HBvBM).  This gives mean filling factors of $\log(\epsilon) = -7.7
\pm 0.1$ for $n_e = 100$cm$^{-3}$, and $\log(\epsilon) = -9.6 \pm 0.1$
for $n_e = 10^3$cm$^{-3}$.
Comparing these filling factors with those in other systems (HBvBM) we
conclude that the most reasonable values for the filling factors are
$\epsilon \sim 10^{-7}$ for $n_e \sim 100$.  We thus estimate that
there is between 0.8\,--\,1.5$\times 10^4$~\msun of ionized
emission-line gas in each of the three elliptical galaxies in this
compact group.

\section{H$\alpha$ Emission Mechanism}
 
In this section we discuss some possible mechanisms that may account
for the H$\alpha$ emission seen in this group.  Despite the fact that
emission line nebulae have been observed in galaxies and cluster
cooling flows for over two decades there is still no general concensus
on the source of excitation that provides both the required energy
flux as well as accounts for the observed spectral-line features (Baum
1992, for a review).

Early observations of isolated early type galaxies indicated that the
\ha luminosity of the emission-line nebulae was correlated with the
X-ray luminosity of the host galaxy (e.g. PJDSB) suggesting that
recombinations in the hot gas could produce the required flux of
emission line gas. 
The \ha luminosity can be estimated under the assumption that
as the gas cools from above $10^6$~K to below $10^3$~K it recombines
and emits (Balmer) photons. The resultant \ha luminosity in HCG62$a$
for a cooling rate $\dot M = 0.5-1$~\msunyr within the
emission line region (estimated from a value of $\dot M =
10$~\msunyr within the cooling radius of 75~kpc and $\dot M \propto r$
[PB]) is $L_{\rm{H}\alpha} \approx 10^{37} H_{\rm rec}$~ergs s$^{-1}$. 
As in other cooling flow systems
(HBvBM, Fabian 1994) each hydrogen atom would have to recombine
between 100-500 times to produce the oberved flux in HCG62$a$ at the
center of the flow.
If the emission line flux was a result of recombination in the
hot gas one might naively expect that at larger distances from the
center, the local cooling and mass accretion rate of the hot gas would
be smaller (since the gas densities are lower) and the galaxies
HCG62$b$ and $c$ would have proportionately smaller fluxes. Since the
observed fluxes are similar the required value of $H_{rec}$ would be
even higher for galaxies further out. Ionization due to recombination
in cooling flow thus fails to account for the observed \ha luminosities,
and photoionization offers a possible mechanism.

A linear correlation exists between the emission-line luminosity and
the absolute B magnitudes of normal early type galaxies (e.g. PJDSB;
Figure \ref{fig:pjdsb}). A recent study by Macchetto et~al.~(1996) also shows that
the emission line flux is even more tightly correlated with the B-band
flux within the line-emitting region. These correlations suggest that
photoionization by stars is a possible mechanism (Johnstone et
al. 1987, HBvBM, McNamara \& O'Connell 1989, Trinchieri \& Di Serego
Alighieri 1991, Allen 1995).
If the gas is photoionized by Lyman continuum photons from stars, the
minimum number of Lyman continuum photons necessary to produce the
observed H$\alpha$ luminosity is
$N_{\rm{Ly}}\,\rm{ph\,s}^{-1}=7.34\times
10^{11}\,L_{\rm{H}\alpha}\,\rm{ergs\,s}^{-1},$ for case B
recombination (Osterbrock 1989). This implies that about $(2-5) \times
10^{50}$~photons s$^{-1}$ are required to explain the current
observations.

Trinchieri \& di Serego Alighieri (1991) suggest photoionization by
post-asymptotic-giant-branch (PAGB) stars
could produce enough ionizing photons to explain the observed
luminosity and satisfactorily account for the observed
spectrum. 
Assuming that each PAGB star emits $5\times 10^{46}$
photons~s$^{-1}$ we would require about $10^4$ stars.  This value is a
lower limit since the covering factor for Lyman radiation is likely to
be less than 1. For a total galactic luminosity $\sim 10^5$~\lsun\
within the line emitting region (3~Kpc), one expects $\sim 0.2$ PAGB
stars at any given epoch (Renzini \& Buzzoni 1986), i.e.\ the chance
of finding one PAGB star in the emitting region is only 20\%. It thus
appears that PAGB stars alone cannot account for the ionization of the
line emitting medium in the HCG62 group galaxies.

We now investigate the possibility of photoionization by young stars.
Spectroscopy of the underlying population in ellipticals by Johnstone
et al.\ (1987) provide evidence for the presence of hot young stars.
If the ionization is due to O stars or OB associations, then 5--10
stars are sufficient to explain the observed H$\alpha$ luminosities,
and the line emitting medium would be in the form of large clumps.
However, narrow band imaging of the line-emitting regions of
ellipticals show a fairly smooth distribution of matter. On the other
hand, intermediate-mass stars could be formed in cool gas providing
the ionizing radiation. The ionizing radiation for an instantaneous
burst of star formation with an IMF of slope 2.5, and a stellar mass
range of 1\msun --30\msun\ is $6\times
10^{45}$~photons~s$^{-1}$~\msun$^{-1}$ (Mayya 1993, 1995). The total
mass in the young stars needed to explain the observations is thus
$(4-8)\times 10^4$\msun. The $e$-folding time for the ionizing
radiation due to evolution of the burst is about $5.5\times
10^6$~yrs. We are hence looking at either a young burst at the present
epoch or a continuous star formation at a rate of (0.007 --
0.015)\msunyr.  The total mass in the young stars estimated here would
require $\ga 10^6$\msun of cold gas assuming star formation efficiency
in normal spirals (0.2-0.5). Cold gas (H$_2$ and \ion{H}{1}) with
masses of the order $10^5-10^6$ have been detected in several early
type galaxies (see Knapp 1990). Star formation with an upper cutoff
$\sim 30$~\msun\ appears to be a viable mechanism to explain the observed
\ha luminosities.

Emission-line nebulae associated with radio galaxies are typically
more luminous than nebulae in normal early-type galaxies.  This has
lead several authors to suggest that active galactic nuclei could
provide a source of ionizing radiation (HBvBM, Shields 1991, B93).
Although there is a correlation between the emission line luminosities
in clusters cooling flow nebulae and radio-power, the variation of the
line ratios as a function of radius are not consistent with a point
source of radiation (HBvBM).  As mentioned earlier the dominant
elliptical galaxy (HCG62$a$) is known to be a weak nuclear radio
source (Menon \& Hickson 1985) which could be an additional source of
ionizing radiation for this galaxy.  It does not, however, account for
the ionized gas in the other two galaxies.

Finally we investigate the importance of turbulent mixing (Begelman \&
Fabian 1990 and Crawford \& Fabian 1992) and the energy input from
buoyancy waves ($g$-waves) generated by galaxy motions through the
intracluster medium (Balbus \& Soker 1990 and Lufkin et al. 1995).  It
is found that neither of these mechanisms satisfactorily account for
the energetics of the emission-line regions, therefore we do not
discuss them in any detail.

Since we find no evidence to suggest that the emission line nebula in
HCG62$a$ is more luminous in a statistical sense than either its
companions or nebulae in typical field elliptical galaxies we are
inclined to believe that the emission-line nebulae in all three
galaxies are sustained by the same mechanism - namely photoionization,
possibly by young intermediate mass stars.  Spectral-line observations
are being planned to determine whether the line ratios support this
mechanism.

\section{Discussion and Summary}

The primary issue arising out of the results presented here is that
despite the fact that the estimated cooling time for the X-ray gas is
much shorter than a Hubble time ($\sim 10^9$~yr), the central dominant
galaxy HCG62$a$ is not more luminous per unit B-band luminosity than
its companions. We emphasise again that while about 60\% of early type
field galaxies have emission line nebulae (B93, Macchetto et
al. 1996), only about 40\% of central cluster galaxies have emission
lines (Edge et al. 1992, Donahue et al. 1992), so the presence of a
cooling flow in X-rays is by no means a sufficient condition for
optical emission lines to be detected.

It is well known (e.g. Forman 1988) that the presence of a dominant
central galaxy is crucial to the existence of a cooling flow. Rich,
X-ray luminous clusters without a dominant galaxy generally have very
irregular X-ray emission and no cooling flow (e.g. Coma cluster)
suggesting that cooling flows are fragile structures that are easily
disrupted by the tidal effect of a large secondary galaxy.  It is
therefore surprising that a cool central component (apparently a
cooling flow) exists in this group where two large galaxies (probably
in an early stage of merger) dominate the central region.  Recent
simulations (for instance by Diaferio et~al.~1995, Pildis,
et~al.~1996) show that substructure within collapsing loose groups
have X-ray and optical properties resembling many compact
groups. These simulations however indicate that mergers are
continuously occuring within the denser regions. This suggests that
the time scale over which the central region is quiescent allowing gas
accretion to occur is very short. The X-ray images of other compact
groups observed to have diffuse X-ray emission are rarely as regular
and centrally concentrated as that associated with the group HCG62. An
indication that the group may have experienced a recent merger comes
from the fact that the outer isophotes of the X-ray halo are not
concentric with the central isophotes (PB), similar to clusters which
have experienced recent mergers.  Recent {\it ROSAT} observations of
NGC~2300 (Davis et al. 1996) indicate that two of the large individual
galaxies in the group are themselves strong X-ray sources. The smooth
image of the group generated by subtracting the point sources
indicates that the diffuse gas emission does not peak on either
galaxy.  It seems likely that higher resolution observations will show
greater structure in the hot gas in the HCG62 group as well, possibly
altering some of the current estimates of the gas accretion rate (Edge
et al. 1992). Alternatively HCG62$b$ could be merely projected on the
center rather than close to it or may have only recently arrived at
the center, explaining the absence of signs of strong tidal
interaction in the optical images. It is possible that once the merger
begins the flow would once again be disrupted as is apparently the
case in HCG94 (Pildis 1995).

  It has been proposed that mergers might actually help to ``inject
kinetic energy into the cluster core'' and power the emission line gas
(Edge et al. 1992, Fabian 1994).  It seems unlikely that the impending
merger in the HCG62 group has enhanced the optical emission fluxes. It
seems likely that in order to be effective in powering emission line
nebulae the merger must be between a galaxy or group of galaxies that
are of late type and which bring with them some cold gas which is then
ionized either by low mass star formation, or thermal conduction from
the hot gas (Macchetto \& Sparks 1992), or by turbulent mixing and
shocks (Crawford \& Fabian 1992). This then suggests that a possible
explanation for why the optical emission-line nebulae in this group
are less luminous than those in cooling flows is that it does not have
a gas rich member and has not recently accreted one. This argument is
supported by the findings of B93  that the emission line regions in
luminous ellipticals are more closely associated with the cold ISM
component than with the hot X-ray component.

To summarise the main findings of this paper:\\ 
We have detected
extended emission line regions at the centers of 3 of the galaxies in
the compact group HCG62. The luminosities in the \hn~ emission line
are of the order $\sim 10^{39}$ \ergss and arise in 3~kpc size regions
in the galaxies HCG62$a$, HCG62$b$, and HCG62$c$.  The luminosities
are comparable to the lower limit of the \hn~-line luminosities
observed in elliptical galaxies. The oberved fluxes imply a total mass
of ionized gas of $5\times~10^3- 10^4$\msun in each of the galaxies.

Although the X-ray cooling flow is centered on the dominant galaxy
HCG62$a$ we find no evidence to suggest that the emission line
luminosity of the central galaxy HCG62$a$ is significantly greater
than that of the other two galaxies in the group leading us to
conclude that these emission line regions resemble those in isolated
ellipticals.

 We have investigated various possible mechanisms for accounting for
the energetics of the ionsized gas regions in the galaxies. We find
that photoionization by starlight, most probably from a population of
young intermediate mass stars can account for the observed emission
line luminosities.

\acknowledgements

We are grateful to the Director and the time allocation committee of
the Indian Institute of Astrophysics for time on the VBO.  We thank
C. J. Jog, R. Pildis, T. P. Prabhu, and J. H. van Gorkom for helpful
discussions and the referee for suggestions for improving this paper.
IRAF is distributed by National Optical Astronomy Observatories which
is operated by the Association of Universities Inc.\ (AURA) under
cooperative agreement with the National Science Foundation.  This
research has made use of the NASA/IPAC Extragalactic Database (NED)
which is operated by the Jet Propulsion Laboratory, Caltech, under
contract with the National Aeronautics and Space Administration.  This
work was begun while both authors were Postdoctoral Fellows at the
Inter-University Centre for Astronomy and Astrophysics (Pune, India).

\end{document}